\def\be{\begin{equation}}
\def\ee{\end{equation}}
\def\bea{\begin{eqnarray*}}
\def\eea{\end{eqnarray*}}
\newcommand{\degree}{\ensuremath{^\circ}}

\documentclass[useAMS,usenatbib]{mn2e}
\usepackage{subfigure}
\usepackage{graphicx}
\usepackage{flafter}
\usepackage{lscape}
\usepackage{longtable}
\begin{document}

\title{AzTEC Millimetre Survey of the COSMOS Field: I. Data Reduction
and Source Catalogue}


\author[K.S. Scott et al.]{
K.S.~Scott,$^1$ 
J.E.~Austermann,$^1$ 
T.A.~Perera,$^1$ 
G.W.~Wilson,$^1$ 
I.~Aretxaga,$^2$
\newauthor
J.J.~Bock,$^3$
D.H.~Hughes,$^2$
Y.~Kang,$^4$ 
S.~Kim,$^4$
P.D.~Mauskopf,$^5$
D.B.~Sanders,$^6$
\newauthor
N.~Scoville,$^7$
and M.S.~Yun$^1$\\
$^1$Department of Astronomy, University of Massachusetts, Amherst, MA 01003.\\
$^2$Instituto Nacional de Astrof\'{i}sica, \'{O}ptica y
Electr\'{o}nica, Tonantzintla, Puebla, M\'{e}xico.\\
$^3$Jet Propulsion Laboratory, California Institute of Technology, Pasadena, CA 91109.\\
$^4$Astronomy \& Space Science Department, Sejong University, Seoul,
South Korea.\\
$^5$Physics and Astronomy, Cardiff University, Wales, UK.\\
$^6$Institute for Astronomy, University of Hawaii, Honolulu, HI 96822\\
$^7$California Institute of Technology, Pasadena, CA 91125.\\
}

\date{\today}

\pagerange{1--16} \pubyear{2008}

\maketitle

\label{firstpage}


\begin{abstract}
We present a 1.1~mm wavelength imaging survey covering 0.3~deg$^2$ 
in the COSMOS field. These data, obtained with the AzTEC continuum
camera on the James Clerk Maxwell Telescope (JCMT), were centred on a
prominent large-scale structure over-density which includes a rich
X-ray cluster at $z \approx 0.73$. A total of 50 millimetre galaxy
candidates, with a significance ranging from 3.5--8.5$\sigma$, are
extracted from the central 0.15~deg$^2$ area which has a uniform
sensitivity of $\sim$1.3~mJy/beam. Sixteen sources are detected with 
$S/N \ge 4.5$, where the expected false-detection rate is zero, of
which a surprisingly large number (9) have intrinsic (de-boosted)
fluxes $\ge 5$~mJy at 1.1~mm. Assuming the emission is dominated by
radiation from dust, heated by a massive population of young,
optically-obscured stars, then these bright AzTEC sources have FIR
luminosities $> 6\times 10^{12}~\rm L_{\odot}$ and star
formation-rates $> 1100~\rm M_{\odot}/yr$. Two of these nine bright
AzTEC sources are found towards the extreme peripheral region of the
X-ray cluster, whilst the remainder are distributed across the
larger-scale over-density. We describe the AzTEC data reduction
pipeline, the source-extraction algorithm, and the characterisation of
the source catalogue, including the completeness, flux de-boosting
correction, false-detection rate and the source
positional uncertainty, through an extensive set of Monte-Carlo
simulations. We conclude with a preliminary comparison, via a stacked
analysis, of the overlapping MIPS 24~\micron~data and radio data with
this AzTEC map of the COSMOS field.
\end{abstract}




\section{INTRODUCTION}
\label{sec:int} 
A decade after the discovery of a population of extremely
luminous, high-redshift dust-obscured galaxies detected by their
sub-millimetre and millimetre wavelength emission
\citep{smail97,hughes98,barger98}, over 200 sub-mm/mm galaxies
(hereafter SMGs) have been detected with signal to noise ratio
$\ge 4$ in blank field surveys
\citep[e.g.,][]{borys03,greve04,laurent05,coppin06} and in surveys
towards moderate redshift clusters designed to probe the faintest SMGs
via lensing \citep[e.g.,][]{smail98,smail02,chapman02}. Their high FIR
luminosities (L$_{FIR}\sim 10^{12-13}~\rm L_\odot$) and inferred
star formation rates \citep[SFR $\gg
100~\rm M_\odot/yr$,][]{smail97,hughes98,barger98} suggest that
these galaxies are high-redshift analogs to the local ULIRG population
\citep{sanders96}, and that they may be the progenitors of the massive
elliptical population observed locally.

Until recently, the relatively modest mapping speeds of SCUBA
\citep[850~\micron,][]{holland99} on the 15-m James Clerk Maxwell
Telescope (JCMT), MAMBO \citep[1.2~mm,][]{kreysa98} on the Institut de
Radio Astronomie Millimetrique (IRAM) 30-m
telescope and Bolocam \citep[1.1~mm,][]{glenn98,haig04} on the 10-m
Caltech Submillimeter Observatory (CSO), have restricted SMG surveys to 
$< 300$~arc-min$^2$ in size, limiting our understanding of
the brightest, rarest SMGs and resulting in wide variations in the
derived number counts as a result of small number statistics and
cosmic variance \citep[e.g.,][]{chapman02,smail02,scott02,borys03}. With new
emphasis on large ($> 300$~arc-min$^2$) sub-mm/mm blank field surveys, 
\citep{greve04,laurent05,mortier05,bertoldi07}, an accurate
characterisation of the bright end of the SMG
number counts and the mean properties of the SMG
population is now becoming possible \citep[e.g.,][]{coppin06}.

We surveyed a 0.15~deg$^2$ region within the COSMOS field
\citep{scoville07a} with uniform sensitivity at 1.l~mm with the AzTEC camera
\citep{wilson08} on the JCMT. The AzTEC survey field is
centred on a prominent large-scale structure as traced by the galaxy
density \citep{scoville07b}, including a massive galaxy cluster at
$z=0.73$ (Figure 1). This AzTEC map has no overlap with the MAMBO/COSMOS survey
\citep{bertoldi07} and only a small amount of overlap with the
shallower Bolocam survey (J. Aguirre, private communication). Both
MAMBO and Bolocam surveys cover a low galaxy-density region of the
COSMOS field, whilst our new AzTEC observations are designed to examine the
impact of massive large-scale foreground structures on SMG surveys in
order to provide a measure of the importance of cosmic variance in the
observed source-density at millimetre wavelengths.

In this paper we present the AzTEC mm survey of the COSMOS field,
including the data reduction and source catalogue. Since this is the
first in a series of papers describing the surveys completed by AzTEC
on the JCMT, we provide an extensive description of the data analysis
pipeline used to extract sources from AzTEC maps.  The JCMT observations,
pointing, and calibration strategy are described in \S~\ref{sec:obs}. A
detailed description of the data reduction algorithm is given in
\S~\ref{sec:dat}.  In \S~\ref{sec:scat}, we present the AzTEC map and
source catalogue, followed by a discussion of simulations used to
determine flux boosting, false-detection rate, completeness, and
source positional uncertainty in the map in \S~\ref{sec:sim}. A
preliminary comparison of the mm sources to the radio and MIPS
24~\micron~populations is made in \S~\ref{sec:multi_compare},
and we discuss the contribution of AzTEC sources to the Cosmic
Infrared Background in \S~\ref{sec:cib}.

The large number of bright SMGs identified in the AzTEC/COSMOS field
strongly suggests a bias in the number density introduced by the known
large-scale structure that is present in the map. A detailed treatment
of this analysis is beyond the scope of this paper and is deferred to
Paper II (Austermann et al., in prep.).  The multi-wavelength imaging
data from the HST/ACS, \textit{Spitzer} IRAC and MIPS, as well as deep
radio imaging from the VLA is particularly valuable for identifying
and studying the nature of the SMGs identified by AzTEC. We will
present a complete study of the multi-wavelength properties of the
SMGs detected in the COSMOS field in Paper III. 

We assume a flat $\Lambda$CDM cosmology with
$\Omega_M = 0.3$, $\Omega_{\Lambda} = 0.7$, and
$H_0 = 73$~km~s$^{-1}$~Mpc$^{-1}$
throughout.

\begin{figure*}
\begin{center}
\includegraphics[width=17.5cm]{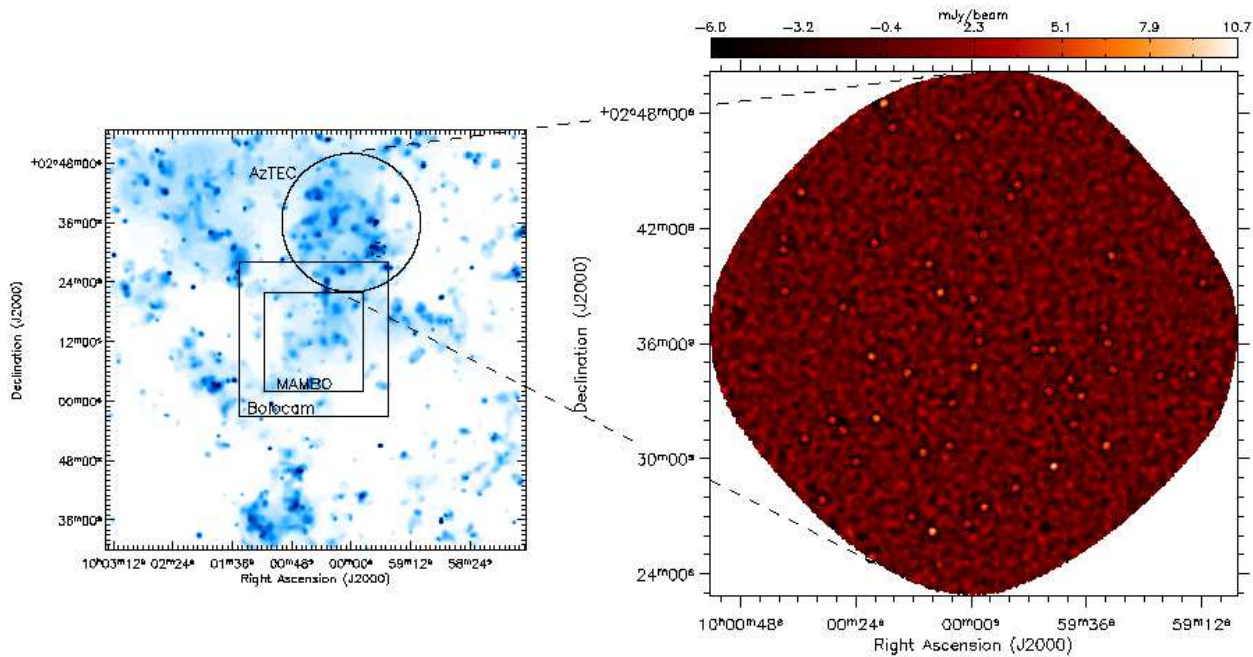}
\caption{{\bf Left:} The galaxy density map from
\citet{scoville07b}, with the boundaries of the AzTEC, Bolocam, and
MAMBO mm surveys within the COSMOS field indicated. The location of
the $z = 0.73$ cluster environment is identified by the dashed
circle. {\bf Right:} The
AzTEC/COSMOS map with $\ge 3.5 \sigma$ source candidates identified by
circles with diameters equal to twice the AzTEC FWHM on the JCMT. The
map has been trimmed to the ``75\% coverage region'' and has an
average rms noise level of 1.3~mJy/beam and an area of
0.15~deg$^2$. The signal map has been Wiener filtered for optimal
identification of sources as described in
\S~\ref{sec:wiener}.}
\label{fig:survey_map} 
\end{center}
\end{figure*}


\section{OBSERVATIONS}
\label{sec:obs}
We selected a 0.3~deg$^2$ region in the northwest quadrant of the
COSMOS field for millimetre imaging with AzTEC. Only the central area
of 0.15~deg$^2$, with uniform noise, is discussed in this paper. The
observations were carried out at the JCMT in November and December
2005. A total of 34 hours of telescope time (excluding pointing and
calibration overheads) was devoted to this survey. 

Details of the AzTEC instrument specifications, performance, and
calibration method at the
JCMT are described in \citet{wilson08} and are briefly summarised
here. The array field of view is roughly circular
with a diameter of 5\arcmin. During the JCMT observing campaign, 107
out of the 144 detectors were operational.
The point spread function (PSF) of the detectors is determined
from beam map observations on bright point sources and is
well described by a two-dimensional Gaussian, with a beam FWHM of
17\arcsec~$\pm$~1\arcsec~ in Azimuth and 18\arcsec~$\pm$~1\arcsec~ in
Elevation.

The COSMOS data-set consists of 34 individual raster-scan observations,
each centred at (RA, DEC) = (10$^{\rm h}$00$^{\rm m}$00$^{\rm s}$, 
+02\degree36\arcmin00.0\arcsec). The observations were taken in
unchopped raster-scan mode by sweeping the telescope in Elevation,
taking a small step of 10\arcsec~in Azimuth, then sweeping back in the
opposite direction, moving only the primary dish. This pattern is
repeated until the entire field
has been mapped.  The small step size ($\sim$1/2 the beam FWHM) and
chosen scan speeds result in a Nyquist-sampled sky with extremely
uniform coverage for each individual observation.

The first half of the observations were taken early in the JCMT
observing run, while scanning strategies were still being
optimised. For these observations, we imaged a 25$\times$25~arc-min$^2$
region, using a scan speed of 90\arcsec/s. From diagnostic tests of
these early AzTEC/JCMT observations, we determined that a faster scan
speed of 150\arcsec/s was optimal, since scanning the camera faster
moves the point-source response to higher temporal frequencies and
away from the low-frequency atmospheric signal, improving the
effectiveness of our cleaning algorithm \citep{wilson08}.  The time
necessary to turn the telescope around between scans (i.e. reverse
direction) is constant and independent of scan speed. Therefore, to
maintain observational efficiency, we expanded the survey region to
30$\times$30~arc-min$^2$ for the later observations

Since the array orientation is fixed in Azimuth and Elevation, the
scan angle in the RA-DEC plane for a raster-scan map
continuously changes due to sky
rotation. When combining several observations with different scan
angles into a single map, we obtain excellent cross-linking that 
suppresses scan-synchronous systematic noise in the maps. We chose to
scan in the Elevation direction rather than in Azimuth to avoid vibrational
noise from the telescope dome motion \citep{wilson08}. 

The opacity at 225~GHz, $\tau_{225}$, was recorded every 10
minutes by the CSO tau monitor. For the AzTEC/COSMOS observations, the
effective opacity, $\tau_{225} \cdot A$, where $A$ is the
airmass, ranged from 0.07--0.27 with an average value of 0.15. The
empirical mapping speed (excluding overheads) derived from the
individual COSMOS observations ranges from
8--34~arc-min$^2$~mJy$^{-2}$~hr$^{-1}$ and is a strong function of
$\tau_{225} \cdot A$ \citep{wilson08}, suggesting that the
noise in each individual observation is dominated by residual
atmosphere that is not removed in the cleaning process. We discuss the
details of atmosphere removal and optimal filtering in the next section.

\begin{figure}
\begin{center}
\includegraphics[width=8cm]{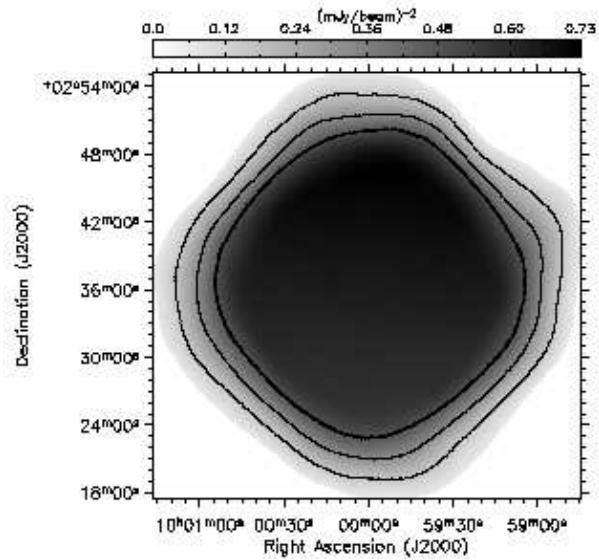}
\caption{The weight map for the AzTEC/COSMOS survey. The contours show
         curves of constant noise and are 1.4, 1.8, and 2.5~mJy/beam
         from the innermost to the outermost contour. The thick,
         innermost contour
         indicates the 0.15~deg$^2$ ``75\% coverage region'' where the
         signal map is trimmed to provide very uniform coverage
         in the region where the analysis in this paper is carried
         out. The noise levels in this central region of the map range from
         1.2 to 1.4~mJy/beam.}
\label{fig:weight_map} 
\end{center}
\end{figure}

\subsection{\label{ssec:pointing}Pointing}
Observations of J1058+015, a variable QSO with a mean flux density of 2.8~Jy,
were made approximately every two hours in order to generate 
small corrections to the JCMT's pointing model.  These corrections
were not made in real time. Instead, a correction based on
a linear interpolation of the measured pointing offsets was applied to
each observation {\it ex post facto}. In \S~\ref{ssec:radio} we
demonstrate that the resulting absolute pointing uncertainty of the AzTEC map
is $<$ 2\arcsec.

\subsection{Flux Calibration}
The AzTEC calibration has been derived from beam map observations of
Uranus, which had a predicted flux density of 44.3--48.5~Jy at 1.1~mm
during the JCMT observing run. We fit a two-dimensional Gaussian to the
PSF of each detector to determine the flux conversion factor (FCF)
from optical loading (in Watts) to source flux (in Jy/beam). Beam maps
were taken once per night. The extinction- and
responsivity-corrected FCF for
each detector did not vary greatly over the entire observing run. 
We use an average FCF for each bolometer determined from all Uranus
beam maps taken at the JCMT. The total error of 6--13\% on the
calibrated signals includes the standard deviation of the measured
FCFs plus errors from the extinction and responsivity corrections
\citep{wilson08}. This value does not
include the 5\% absolute uncertainty in the flux density of Uranus
\citep{griffin93}. The data are calibrated after atmosphere removal
and before combining the time-stream signals from all bolometers into
a single map.


\section{DATA REDUCTION}
\label{sec:dat}
The AzTEC/COSMOS data-set is reduced using the publicly available
AzTEC Data Reduction Pipeline V1.0 written in IDL and developed by
AzTEC instrument team members at the University of Massachusetts,
Amherst. V1.0 has been optimised for the identification of point
sources in blank-field extragalactic surveys.  The 34 individual
raster-scan observations that comprise the AzTEC/COSMOS data-set are ultimately
combined to produce four data products: 1) a co-added signal map; 2) a
corresponding weight map; 3) a set of noise maps which are
representative of the noise in the co-added signal map; and  4) a
representation
of the instrument point source response, post-cleaning and filtering.
We describe the techniques for creating these data
products from raw AzTEC data in detail in this section.

The raw data-file for each raster-scan observation is composed of bolometer
signals, telescope pointing signals, and environmental signals -- all
stored as a function of time and referred to hereafter as
``time-stream'' data.  Detector signals are sampled at a rate of 64~Hz
and all germane environmental signals are interpolated to this sampling
rate in the analysis.  In the description below, a ``scan'' is defined
as a single constant-velocity and constant-Elevation pass of the telescope from
one side of the field to the other. We do not use the data recorded as
the telescope is strongly accelerating at the ends of the scans
(during the turn-around), where the
accuracy of the pointing signals is unknown and micro-phonic noise is
more likely. Given  the field size and scan velocities used for the
AzTEC/COSMOS survey, this results in a loss of 22--34\% of the on-source
observing time.

\subsection{De-spiking}
Prior to atmosphere removal, the data are inspected for cosmic ray
events and instrumental glitches, both of which register as ``spikes''
in the raw time-stream data. Spikes in the AzTEC data occur at a rate
of $\sim$40~hr$^{-1}$, each usually confined to a single detector, and
with amplitudes that vary widely from 30~mJy
to 550~Jy.  Spikes are defined in our automated spike identification and
removal procedure as any instance where a detector signal jumps by a
user-defined threshold (typically $> 7\sigma$ or $< 7\sigma$) between
adjacent time
samples. Generally, such jumps in detector output cannot be of
astronomical origin as the continuous nature of the beam and the scanning 
strategy ensure a smoother signal.  Spikes are located recursively, thus
allowing for pairs of spikes with high dynamic range to be identified
independently.  A spike decay length (time necessary for the spike signal to 
drop below the baseline noise rms) is calculated based on the spike 
amplitude and a conservative estimate of the detector time constant.  
Adjacent samples are flagged accordingly, with a 
minimum of 12 (6) samples flagged after (before) the spike.
Flagged data samples are not included in the map-making process.
For the AzTEC/COSMOS data-set, flagged samples due to spikes account
for $<$ 0.1\% of the total time-stream data.

Since the matrix operations in our atmosphere removal technique
requires that all bolometers
have the same number of time-stream samples, we cannot simply discard
the flagged samples. Large spikes can affect upwards of $\sim$20 adjacent time
samples for a single detector and de-correlate that detector's time
stream from the remainder of the array.  Unaccounted for, this would
reduce the efficacy of the atmospheric cleaning technique and so
we replace each set of flagged samples with the sum of two components:
1) Gaussian noise with variance equal to the variance of that
detector's time-stream from nearby unflagged samples; and 2) an
appropriately scaled baseline calculated from the mean time-stream for
all unaffected 
detectors.  In this manner, the detector-detector covariance matrix is
minimally affected and, more importantly, the inclusion of noise
ensures that excess weight is not given to the synthetic time-stream
samples.  These simulated data are used \textit{only} 
in the atmosphere removal process; all flagged samples are discarded
when making the actual map.

\subsection{\label{sec:atmosphere_removal}Atmosphere Removal}
The signal due to the fluctuating atmosphere dominates the background
SMG population by three orders of magnitude.  For the
AzTEC/COSMOS data-set and other ``blank-field'' surveys we adopt an
adaptive principal component analysis (PCA) technique 
similar to that described by \citet{laurent05} to
remove, or ``clean'' the correlated sky noise from the time-stream data.  Faint
point sources are, in general, not correlated between detectors in the
array while the atmosphere is correlated on all spatial scales of
interest.  The adaptive PCA technique uses the degree of correlations
to distinguish between the two. 

Cleaning is accomplished on a scan by scan basis.  The basic adaptive
PCA cleaning process is as follows: a covariance matrix is constructed
from the $N_{\rm{bolo}}$ by $N_{\rm{time}}$ de-spiked time-stream
data for each scan and then eigenvalue decomposed.  The relative
amplitudes of the resulting eigenvalues are representative of the
degree of correlation of the detector signals for the mode described
by the respective eigenvector.  Since fundamental detector noise and
faint point sources are not correlated amongst multiple detectors,
they will not lie preferentially in modes having large eigenvalues. 
The atmosphere, fluctuations in the detector bias chain, 
and other common-mode signals dominate the correlated variance
with their power in modes with large eigenvalues. The technique, then,
is to identify and project out modes with the largest eigenvalues.

\begin{figure*}
\includegraphics[width=16cm]{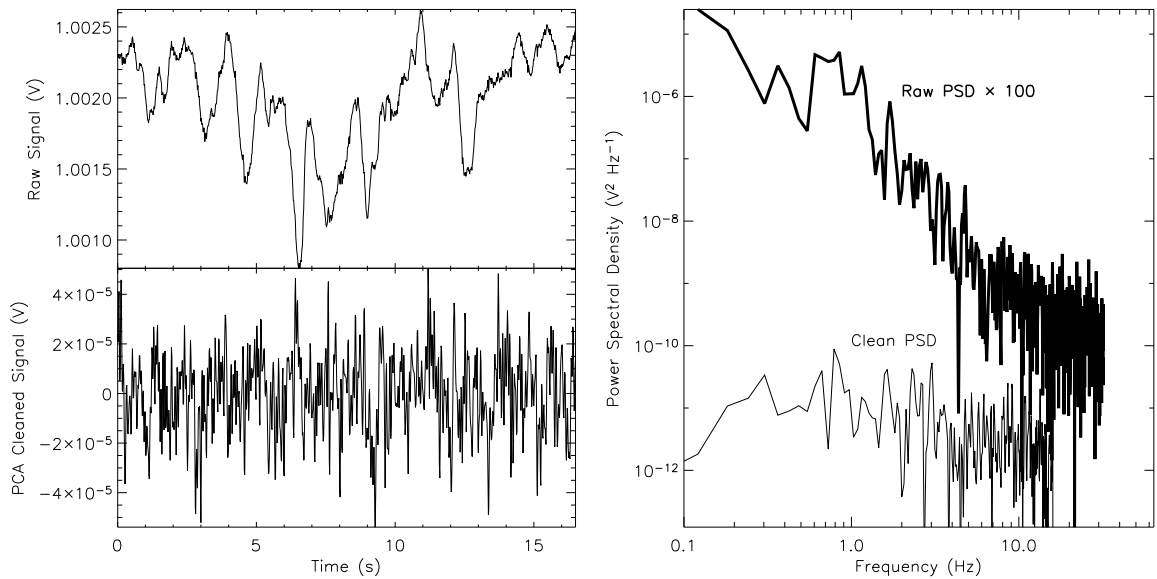}
\caption{{\bf Top Left:} The raw time-stream signals for a sample
         bolometer during a single scan. {\bf Bottom Left:} The same
         time-stream signals after PCA cleaning. Note the factor of 20
         reduction in the noise level post-cleaning. {\bf Right:} The power
         spectral density (PSD) of the same scan, before (thick)
         and after (thin) PCA cleaning, demonstrating the reduction of
         low-frequency signal. The PSD before PCA cleaning has been
         multiplied by a factor of 100 to offset the two curves. The
         PSD of the post-cleaned
         data is truncated at 16~Hz due to a digital low-pass filter
         that is applied to the data before PCA cleaning.}
\label{fig:pca} 
\end{figure*}

The choice of which modes to remove from the data is somewhat
arbitrary.  Empirically
we have found the following to work well.  First, the mean and
standard deviation in the base-10 logarithm of the eigenvalue
distribution is determined, then
large eigenvalues that are $> 2.5\sigma$ from the mean are cut. This process is
repeated until no $> 2.5\sigma$ outliers exist. An example of the time
stream data and power spectral density (PSD) before and after PCA
cleaning is shown in Figure \ref{fig:pca}. The significant decrease in
the power at low frequencies demonstrates how this adaptive PCA cleaning
technique effectively removes much of the atmospheric signal.

There are two consequences of the adaptive PCA technique that must be
addressed.  First, since faint point sources have power at low spatial
frequencies, there is no way to completely decouple the atmosphere
from the point source signal. We therefore expect some attenuation of
point sources in the resulting map. Secondly, PCA cleaning AC-couples
the time-stream
signal, leaving the mean of the samples for each bolometer in a single
scan equal to zero.

We trace the effects of PCA cleaning on the point source response
profile and its amplitude to generate the point source kernel, which
we use later in the analysis to optimally filter the map and correct for
the attenuation. Since the degree of 
attenuation varies according to the conditions of the atmosphere for
a given observation, we create a point source kernel for each
observation separately. The procedure is as follows: 1) each scan of an
observation is cleaned according to the prescription given above,
saving the set of eigenvalues and eigenvectors for later use;  2) an
analogous, synthetic time-stream is created using the pointing
signals to make a fake ``observation'' of a 1~Jy point source centred
in an otherwise empty and noiseless field. The flux of the synthetic
point source is arbitrary -- we only need to determine the factor of
attenuation and the effect that PCA cleaning has on the shape of the
point source response;  3) the dominant
eigenvectors identified in 1) are projected out from the synthetic
data; and  4) a map is made from this cleaned,
synthetic data.  The resulting image is the point source kernel, and it
has the same shape and attenuation as a point source in the cleaned
signal map for a given observation.  This is
true only if the real sources in the time-stream signal do not
significantly affect the PCA cleaning, and if the kernel does not vary
greatly in shape and attenuation across the whole
field. The standard deviation and spatial PSD of an individual signal
map is comparable to that in a jackknifed noise realisation of that map (see
\S~\ref{ssec:noisemap}), which suggests that the former must be true. 
We have tested the latter assumption 
by placing the synthetic 1~Jy point source at different locations in
the field. We find that the shape of the kernel is not affected by its
location, and the measured peak of the PCA-cleaned kernel varies by
less than 2\% over the entire field.

\begin{figure}
\includegraphics[width=8cm]{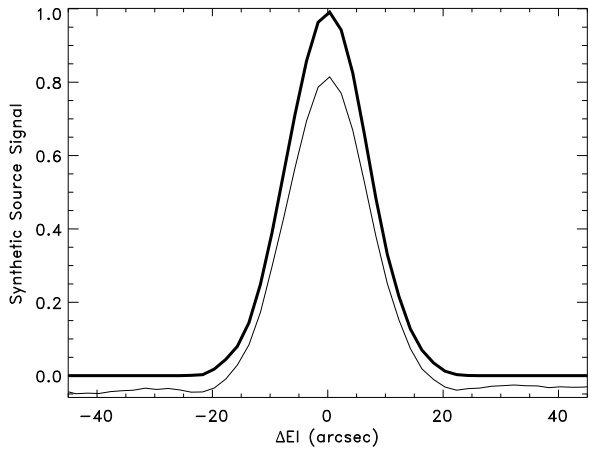}
\caption{A cut in Elevation of the point source kernel for an
individual observation. The
thick curve shows the effective PSF (once all
bolometer signals are combined) \textit{before} PCA cleaning. The thin
curve shows the resulting point source response function after the
synthetic source has been PCA cleaned in the same manner as the real
time-stream signals.}
\label{fig:kernel}
\end{figure}

In Figure~\ref{fig:kernel}, we show a cut in Elevation 
through the synthetic point source for one of the observations,
before and after PCA
cleaning. This demonstrates the attenuation that a real source 
experiences from the atmosphere removal process. In this case, the
sources will be attenuated by 17.8\% due to PCA cleaning. This also
shows how the cleaning affects the shape of point sources. The central
peak is now flanked with negative side-lobes and has a small negative
baseline that extends across the map, making the mean of the point
source response equal to zero.

\subsection{Raw Signal Maps}
We cast each of the 34 individual raster-scan observations into map
space prior to co-adding them into a single map. Hereafter, we will
refer to any maps that are made for a single observation as an
``individual'' map. To ensure that all of these
individual maps will have the same coordinate grid, we convert the
time-stream 
pointing signals into offset positions relative to the map centre at
(RA, DEC) = (10$^{\rm h}$00$^{\rm m}$00$^{\rm s}$, 
+02\degree36\arcmin00\arcsec). These pointing signals are then binned into
$2\times2$~arc-sec$^2$ pixels, creating the underlying coordinate grid
for the map.
We chose 2\arcsec~pixelization in order to avoid significant dilution
of the peak signal from point sources while maintaining a
statistically sufficient number of samples ($\ge 9$) in each
pixel. The map value for pixel $j$ in observation $i$, $S_{i,j}$, is
calculated from
the weighted average of all samples whose central pointing falls within the
pixel boundary, combining the samples from all bolometers
simultaneously and excluding any samples flagged in the de-spiking
process.  The weight of each sample is taken to be the inverse
variance of the respective detector's samples in the parent
scan. This weighting scheme is only suitable for cases where the
source signal is consistent with noise for a single scan observation,
which is true for the entire AzTEC/COSMOS data-set.

For each individual COSMOS map, $S_i$, we also make the corresponding
individual ``weight map'', $W_i$, by adding in quadrature the weights of all
bolometer samples that contribute to a pixel.  As the flux assigned to
a pixel is a weighted average of these samples, the weight of a pixel
is proportional to $\sigma_{i}^{-2}$ of the flux estimate.  The
proportionality constant may differ from unity because all samples
contributing to a pixel may not be completely independent, for
instance due to detector-detector correlations resulting from
imperfect atmosphere removal.  However, because the scan strategy and
analysis technique are essentially identical for all observations, we
expect on average that this proportionality constant is identical over
the 34 individual observations and over all pixels of an individual
map.  As noted before we also make an image of the point source kernel,
$K_i$, for each individual observation.

We combine all individual COSMOS observations into a single image by computing
for each pixel the weighted average over the individual maps:
\be
S = \frac{\sum_{i=1}^{34} W_iS_i}{\sum_{i=1}^{34} W_i}.
\ee
As with each of the individual observations, we also produce the 
weight map, $W$, corresponding to this co-added signal map and an averaged
point source kernel, $K$.

\subsection{\label{ssec:noisemap}Noise Maps}
With the construction of $S$, $W$, and $K$ we have most of the raw
ingredients for making the final map.  In order to optimally
filter $S$, however, we must construct an estimate of the noise in
$S$.  We do this by generating ``jackknifed'' noise realisations for
each COSMOS observation.  This is accomplished by multiplying
each scan in the cleaned time-stream data by $\pm$1 (chosen at
random) before the map-making process.  This removes the sources, both
resolved and confused, from the bolometers' signals while preserving the noise
properties in the individual scans. We then combine jackknifed noise
realisations made from each of the 34 observations in the same manner as for
the real individual maps to create a single co-added noise map, $N$. 
We choose to jackknife on single-scan scales to ensure a statistically
significant number of elements (there are 150--200 scans per observation) and
to ensure nearly equal weightings in the positive and negative components
while conserving low-frequency components (each scan is $>$ 10 seconds
and $\ge$ 25\arcmin~in length).
This was tested against the more traditional approach to jackknifing, where
half the original individual signal maps are multiplied by a factor of
-1 before combining 
the full data-set, which gave consistent results.

For the AzTEC/COSMOS data-set we create five jackknifed noise realisations for
each of the 34 COSMOS observations. To verify that these noise realisations
are consistent with the noise in the individual signal maps, we compare the
standard deviation and the spatial PSD of the noise realisations to
those in the raw individual 
signal maps directly. This test is valid since the contribution from real
sources in the individual signal map for a single observation is
negligible. We find that the difference between the standard deviations
of the individual signal maps and their jackknifed noise realisations
is less than
0.6\% for every observation. We use random combinations of
these noise realisations, one representing each individual observation
at a time, to generate a
total of 100 co-added noise maps for the field - each a
realisation of the underlying noise in the co-added signal map, $S$.  As
described below, these noise maps are used in creating the
optimal point source filter for the co-added signal map, and as the
underlying noise in synthetic source maps.

\subsection{\label{sec:wiener}Optimal Filtering}

At this stage in the analysis, pixel-to-pixel signal variations stand
out prominently in the co-added signal map.  These variations are not of
astronomical origin as the pixel size, 2\arcsec, is much smaller than
the AzTEC beam.  One way to filter out such features is to convolve
the signal map with our co-added point source kernel, $K$.  The
resulting map must then be scaled to account for attenuation of the
kernel from PCA cleaning.  If the noise covariance matrix of the
signal map were diagonal, that is, if the errors in the pixel values
were independent, then this two-step procedure would be mathematically
equivalent to a fitting procedure: that of shifting the centre of $K$
to the centre of each pixel in $S$ and fitting it to the signal
map to find a best-fit amplitude.  The $K$-convolved scaled map is
equivalent to a map of those best-fit amplitudes.  This analogy to
fitting is useful since it provides guidance on generalising the
filter/convolution procedure and on propagating the error/weight map.

The presence of excess
long wavelength noise in the Fourier transform of noise maps
is clear evidence of pixel-pixel noise correlations.  We de-weight
these long wavelength modes by filtering the signal map with the
inverse of the square root of the power spectral density, averaged
over the 100 noise maps.  This filter makes the noise power flat with
frequency or, equivalently, removes pixel-pixel correlations in the
filtered map.  This ``whitening'' filter is applied to both the
signal map and the point source kernel.  At this point, a linear 
convolution of the two is the same as fitting the whitened kernel to the
whitened map {\em assuming a uniform uncertainty for all pixel
values.}  Such a fit/convolution is equivalent to the conventional
``optimal filtering'' procedure used by other groups
\citep[e.g.,][]{laurent05}, but we follow the fit analog to completion by
including non-uniform coverage as non-constant error values in the fit.

The proper accounting of non-uniform coverage is important for two reasons.
First, implicit to such map-making and
filtering procedures is the assumption that the sky as seen by AzTEC
can be described by a set of discrete points - the centres of the map
pixels.  For large pixel sizes, this assumption is invalid and results
in fluxes (e.g. from point sources) being smoothed out.  Therefore, we
would like to explore the use of small pixel sizes.  While raster-scan
maps made
with AzTEC have rather uniform coverage on beam scales, the coverage
has non-uniformity on small scales like 2\arcsec.  Some groups
\citep[e.g.,][]{coppin06} seek an ``optimal'' pixel size that is small
enough to avoid
flux-smoothing effects and large enough for the coverage variations
between pixels to be negligible.  But such an optimum may not exist.
By including variations in coverage as variable error values in a
fitting procedure, we circumvent having a lower limit to the pixel
size, save for practical CPU time considerations.  Empirically, we
have found that pixel sizes below 3\arcsec~yield essentially the same
results in terms of fluxes and sources recovered in AzTEC/JCMT maps.

Second, the error values are formed from our estimate of the uncertainty
of each pixel value.  Thus, our estimate of the sky coverage of each
pixel is correctly propagated through the analysis, resulting in a new
weight map that represents the formal weight in the best-fit amplitudes at each
pixel.  In summary, the optimal filter consists of 1) finding the
best-fit amplitude from fitting a whitened point-source kernel to
every pixel of a whitened signal map with proper account for the
uncertainty of each pixel value, and 2) propagating the weights
to yield a new weight map representing the uncertainty in the best-fit
amplitude at each pixel.  The signal map times the square root of this
weight map represents the signal to noise for each pixel.

The above filtering procedure is implemented with linear convolutions,
made quicker by the use of fast Fourier transforms.  In the optimal filter, a
rotationally symmetrized version of the point-source kernel is used. 
This is a better approximation to point sources over the entire map
than the raw kernel averaged over individual observations, which has
scan-oriented artifacts that are relevant only to a particular central region
of the map.  We also make use of noise maps to avoid lengthy
calculations and to find an absolute normalisation factor for values
in the final weight map.  The mathematical formulation of this optimal
filter and the details of its implementation will be presented in a
future work.

\section{SOURCE CATALOGUE}
\label{sec:scat}
The AzTEC/COSMOS signal map and its weight map are shown in
Figures~\ref{fig:survey_map} and \ref{fig:weight_map}. The signal map
shown has been trimmed such that only pixels with weights $\ge$ 75\% of
the map's characteristic (roughly the maximum) weight are
included. This results in a nearly circular map with total area
0.15~deg$^2$ and very uniform noise across the map, ranging from
1.2~mJy/beam in the centre to 1.4~mJy/beam at the edges of the
map. Unless 
otherwise  stated, we limit our analysis to this ``75\% uniform
coverage region''.

\begin{figure}
\includegraphics[width=8cm]{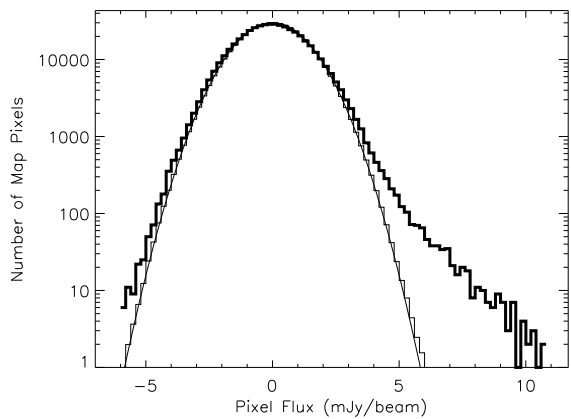}
\caption{Histogram of fluxes from the signal map (thick line)
         and the average histogram of fluxes from the noise maps
         (thin line) with the best-fit Gaussian over-plotted.  A clear
         distortion of the map pixel flux values from that expected
         from noise is seen in the signal map due to the presence of
         real sources.}
\label{fig:flux_hist}
\end{figure}

Figure~\ref{fig:flux_hist} shows the histogram of the pixel
flux density values in the map. The averaged histogram of pixel values from
the filtered noise maps, which is well-fit by a Gaussian with
$\sigma = 1.3$~mJy/beam, is also shown for comparison. There is a
clear excess of positive flux pixels in the signal map compared to the
noise maps, indicating the presence of both bright and confused
sources. The presence of real sources in the map also
produces an excess of hot \textit{negative} flux pixels over that
expected from Gaussian random noise due to the fact that our map is
AC-coupled with a mean of zero. Each source in the map is a scaled version
of the point source kernel and contributes excess negative signal due
to the negative side-lobes surrounding the central peak (see
Figure~\ref{fig:kernel}). Real
sources change the distribution of flux values
in the map from that expected of pure Gaussian noise by skewing
the flux distribution (making it very non-Gaussian), broadening the
distribution, and shifting the peak to $< 0$.

Bright source candidates are identified in the signal to noise map as
local maxima within an 18\arcsec~window above a signal to noise
($S/N$) threshold of 3.5. We find that reducing the
``single-source'' window from 18\arcsec~to 4\arcsec~results in the
same number of source detections. While none of these sources are visually
extended, it is possible that some of our
individually-detected sources consists of multiple components blending
together due to the large beam of the instrument. We could attempt to
``de-blend'' detected sources by fitting them to a combination of two
(or more) point source kernels, but this is precluded by the low
signal to noise of the detections that makes it difficult to
distinguish between a single source versus multiple blended sources.
Sub-pixel centroiding of the source coordinates is calculated by
weighting the pixel positions within a 9\arcsec~radius of the
brightest pixel by the flux squared. This method
results in a list of 50 source candidates with $S/N \ge 3.5$,
which are listed in Table~\ref{table:sources}. The measured
flux density for a source is given by the map value at its peak,
and the error on the flux density by the noise in that pixel. Note
that the optimal filter correctly scales the flux values in the map to account
for the flux attenuation arising from PCA cleaning. The 
``de-boosted'' 1.1~mm fluxes for the AzTEC/COSMOS source candidates listed in
Table~\ref{table:sources} represent the maximum likelihood flux
density using the semi-Bayesian approach outlined in the
following section.

We find a large number of very bright, high-significance
sources in our map, 9 of which have intrinsic fluxes 
$\ge$ 5~mJy. Assuming a modified blackbody spectral energy
distribution (SED) with dust temperature $T_d = 40$~K and
emissivity $\beta = 1.6$, these very bright AzTEC galaxies
have L$_{FIR} > 6.0\times10^{12}~\rm L_{\odot}$. Assuming
that all of the bolometric
output arises from star formation and the relationship between SFR
and L$_{FIR}$ for starburst galaxies from \citet{kennicutt98},
this implies SFRs $> 1100~\rm M_{\odot}/yr$. Seven of these
sources have been followed-up with interferometric imaging at
890~\micron~using the Submillimeter Array (SMA) \citep{younger07}. All
of these sources were detected with the SMA with signal to
noise $\ge 6$ (see Table~\ref{table:sources}),
confirming the reality of these sources
and providing 0.2\arcsec positional accuracy. With the
2\arcsec~resolution of the SMA, none of these seven SMGs were resolved
into multiple components, implying physical sizes of
$< 16$~kpc at $z = 2.2$ \citep[the median
redshift of SMGs from][]{chapman05} and
$< 13$~kpc at $z > 4$, where a fraction of these
SMGs are likely to exist based on their faintness or non-detection in
the radio \citep{younger07}.

From the 1.1~mm number counts of \citet{laurent05}, we expect
on average only 
4--5 sources with intrinsic flux density $\ge$ 5~mJy in a blank,
unbiased field of this size, compared to the 9 discovered in the
AzTEC/COSMOS map. Our map deliberately surveys
a biased portion of the COSMOS field
(Figure~\ref{fig:survey_map}) by being centred on prominent
large-scale structure as traced by the galaxy density map of
\citet{scoville07b}, and there is evidence for a correlation between
the positions of the SMGs in the AzTEC map and the projected galaxy
density for galaxies with $z \le 1.1$ (Austermann et al., in prep. --
Paper~II).  However, for all seven SMGs detected with the SMA, optical
and/or radio/far-IR photometric redshifts place the sources behind the
foreground structure at $z = 0.73$ \citep{younger07}. If some
or all of the $\ge 5$~mJy sources are lensed, then the
bolometric luminosity and SFR calculated above could be significantly
overestimated. In Paper II, we
will present a complete analysis of the relationship between the SMG
population and the foreground galaxy population, including number
counts derived from this study as compared with those from known
blank-fields, a study of possible galaxy-galaxy lensing of the bright
AzTEC/COSMOS sources due to the foreground structure, and several
quantitative tests of the correlation of the AzTEC sources with the
projected galaxy over-density and weak-lensing mass maps.

\begin{table*}
\caption{\label{table:sources}AzTEC/COSMOS source
candidates. The columns give: 1) AzTEC source name; 2) SMA
identification; 3) Signal to noise of the detection in the AzTEC map;
4) Measured 1.1~mm flux density and error; 5) De-boosted flux density
and 68\% confidence interval (\S~\ref{ssec:bayes});
6) 890~\micron~flux density and error 
\citep{younger07}; and 7) Probability that the source will de-boost
to $< 0$ (\S~\ref{ssec:bayes}).}
\begin{center}
\begin{tabular}{llccccc}
\hline
\hline
                          &            &       & $S_{1.1mm}$  & $S_{1.1mm}$         &                     &                      \\
                          &            &       & (measured)   & (de-boosted)        & $S_{890\micron}$    &                      \\
Source                    & SMA ID     & $S/N$ & (mJy)        & (mJy)               & (mJy)               & $P(S_{1.1mm}<0)$     \\
\hline
AzTEC\_J095942.68+022936.0 & AzTEC1 & 8.3 & $10.7\pm1.3$ & $9.3^{+1.3}_{-1.3}$ & $15.6\pm1.1$ & 0.000 \\
AzTEC\_J100008.03+022612.1$^{a,b}$ & AzTEC2 & 7.4 & $ 9.7\pm1.3$ & $8.3^{+1.3}_{-1.3}$ & $12.4\pm1.0$ & 0.000 \\
AzTEC\_J100018.25+024830.2$^{b,c}$ & AzTEC7 & 6.4 & $ 8.8\pm1.4$ & $7.1^{+1.4}_{-1.4}$ & $12.0\pm1.5$ & 0.000 \\
AzTEC\_J100006.40+023839.8 & AzTEC6 & 6.3 & $ 7.7\pm1.2$ & $6.3^{+1.3}_{-1.2}$ & $8.6\pm1.3$ & 0.000 \\
AzTEC\_J100019.73+023206.0$^{b,c}$ & AzTEC5 & 6.2 & $ 7.9\pm1.3$ & $6.5^{+1.2}_{-1.4}$ & $9.3\pm1.3$ & 0.000 \\
AzTEC\_J100020.71+023518.2$^b$ & AzTEC3 & 5.9 & $ 7.4\pm1.2$ & $5.9^{+1.3}_{-1.3}$ & $8.7\pm1.5$ & 0.000 \\
AzTEC\_J095959.33+023445.8$^{b,c}$ & & 5.7 & $ 7.1\pm1.2$ & $5.5^{+1.3}_{-1.3}$ & & 0.000 \\
AzTEC\_J095957.22+022729.3$^{a,e}$ & & 5.6 & $ 7.2\pm1.3$ & $5.8^{+1.3}_{-1.5}$ & & 0.000 \\
AzTEC\_J095931.83+023040.2 & AzTEC4 & 5.3 & $ 6.7\pm1.3$ & $5.2^{+1.3}_{-1.4}$ & $14.4\pm1.9$ & 0.001 \\
AzTEC\_J095930.77+024034.2$^b$ & & 5.1 & $ 6.2\pm1.2$ & $4.7^{+1.3}_{-1.3}$ & & 0.001 \\
AzTEC\_J100008.80+024008.0$^{b,c}$ & & 5.1 & $ 6.2\pm1.2$ & $4.7^{+1.3}_{-1.3}$ & & 0.001 \\
AzTEC\_J100035.37+024352.3$^{b,c}$ & & 4.8 & $ 6.1\pm1.3$ & $4.5^{+1.3}_{-1.5}$ & & 0.003 \\
AzTEC\_J095937.04+023315.4$^{b,c}$ & & 4.8 & $ 6.0\pm1.3$ & $4.4^{+1.3}_{-1.4}$ & & 0.003 \\
AzTEC\_J100010.00+023020.0 & & 4.7 & $ 6.0\pm1.3$ & $4.3^{+1.4}_{-1.4}$ & & 0.005 \\
AzTEC\_J100013.21+023428.2$^b$ & & 4.6 & $ 5.8\pm1.3$ & $4.2^{+1.3}_{-1.4}$ & & 0.005 \\
AzTEC\_J095950.29+024416.1 & & 4.5 & $ 5.4\pm1.2$ & $3.9^{+1.3}_{-1.3}$ & & 0.006 \\
AzTEC\_J095939.30+023408.0$^{b,c}$ & & 4.4 & $ 5.4\pm1.2$ & $3.8^{+1.4}_{-1.4}$ & & 0.011 \\
AzTEC\_J095943.04+023540.2 & & 4.3 & $ 5.4\pm1.2$ & $3.8^{+1.3}_{-1.5}$ & & 0.012 \\
AzTEC\_J100028.94+023200.3$^{b,c}$ & & 4.3 & $ 5.4\pm1.3$ & $3.8^{+1.3}_{-1.6}$ & & 0.016 \\
AzTEC\_J100020.14+024116.0$^{b,c}$ & & 4.3 & $ 5.2\pm1.2$ & $3.6^{+1.3}_{-1.4}$ & & 0.014 \\
AzTEC\_J100002.74+024645.0$^b$ & & 4.2 & $ 4.9\pm1.2$ & $3.4^{+1.3}_{-1.4}$ & & 0.016 \\
AzTEC\_J095950.69+022829.5$^{b,c}$ & & 4.2 & $ 5.4\pm1.3$ & $3.6^{+1.5}_{-1.6}$ & & 0.022 \\
AzTEC\_J095931.57+023601.5$^b$ & & 4.1 & $ 5.1\pm1.2$ & $3.4^{+1.4}_{-1.5}$ & & 0.021 \\
AzTEC\_J100038.72+023843.8$^{b,c}$ & & 4.1 & $ 5.0\pm1.2$ & $3.3^{+1.4}_{-1.5}$ & & 0.024 \\
AzTEC\_J095950.41+024758.3$^b$ & & 4.1 & $ 4.9\pm1.2$ & $3.3^{+1.4}_{-1.4}$ & & 0.024 \\
AzTEC\_J095959.59+023818.5 & & 4.0 & $ 5.0\pm1.2$ & $3.3^{+1.4}_{-1.5}$ & & 0.027 \\
AzTEC\_J100039.12+024052.5$^b$ & & 4.0 & $ 5.0\pm1.2$ & $3.3^{+1.4}_{-1.6}$ & & 0.028 \\
AzTEC\_J100004.54+023040.1$^{b,c}$ & & 4.0 & $ 5.1\pm1.3$ & $3.3^{+1.5}_{-1.6}$ & & 0.035 \\
AzTEC\_J100026.68+023753.7 & & 4.0 & $ 4.9\pm1.2$ & $3.3^{+1.4}_{-1.6}$ & & 0.032 \\
AzTEC\_J100003.95+023253.8 & & 4.0 & $ 5.0\pm1.3$ & $3.3^{+1.4}_{-1.6}$ & & 0.036 \\
AzTEC\_J100034.59+023102.0 & & 3.9 & $ 5.0\pm1.3$ & $3.1^{+1.5}_{-1.6}$ & & 0.040 \\
AzTEC\_J100020.66+022452.8$^b$ & & 3.8 & $ 5.4\pm1.4$ & $3.1^{+1.7}_{-2.0}$ & & 0.071 \\
AzTEC\_J095911.76+023909.5 & & 3.8 & $ 5.0\pm1.3$ & $3.0^{+1.6}_{-1.8}$ & & 0.060 \\
AzTEC\_J095946.66+023541.9$^{b,c}$ & & 3.7 & $ 4.6\pm1.2$ & $2.8^{+1.5}_{-1.7}$ & & 0.056 \\
AzTEC\_J100026.68+023128.1 & & 3.7 & $ 4.8\pm1.3$ & $2.8^{+1.6}_{-1.7}$ & & 0.061 \\
AzTEC\_J095913.99+023424.0 & & 3.7 & $ 4.7\pm1.3$ & $2.8^{+1.5}_{-1.7}$ & & 0.060 \\
AzTEC\_J100016.31+024715.8 & & 3.7 & $ 4.6\pm1.3$ & $2.7^{+1.5}_{-1.8}$ & & 0.067 \\
AzTEC\_J095951.72+024337.9$^{b,c}$ & & 3.7 & $ 4.4\pm1.2$ & $2.6^{+1.5}_{-1.6}$ & & 0.060 \\
AzTEC\_J095958.28+023608.2$^b$ & & 3.6 & $ 4.5\pm1.2$ & $2.7^{+1.5}_{-1.8}$ & & 0.069 \\
AzTEC\_J100031.06+022751.5$^b$ & & 3.6 & $ 4.9\pm1.3$ & $2.7^{+1.6}_{-2.1}$ & & 0.086 \\
AzTEC\_J095957.32+024141.4$^b$ & & 3.6 & $ 4.4\pm1.2$ & $2.6^{+1.4}_{-1.7}$ & & 0.068 \\
AzTEC\_J095930.47+023438.2$^{b,c}$ & & 3.6 & $ 4.5\pm1.2$ & $2.6^{+1.5}_{-1.8}$ & & 0.074 \\
AzTEC\_J100023.98+022950.0 & & 3.6 & $ 4.6\pm1.3$ & $2.6^{+1.5}_{-1.9}$ & & 0.080 \\
AzTEC\_J095920.64+023416.7$^b$ & & 3.6 & $ 4.5\pm1.2$ & $2.6^{+1.5}_{-1.8}$ & & 0.077 \\
AzTEC\_J095932.26+023649.5$^b$ & & 3.6 & $ 4.4\pm1.2$ & $2.6^{+1.4}_{-1.8}$ & & 0.075 \\
AzTEC\_J100000.79+022636.0 & & 3.6 & $ 4.6\pm1.3$ & $2.6^{+1.5}_{-2.0}$ & & 0.088 \\
AzTEC\_J095938.63+023146.2$^b$ & & 3.6 & $ 4.5\pm1.3$ & $2.6^{+1.5}_{-1.9}$ & & 0.086 \\
AzTEC\_J095943.74+023329.9$^{b,c}$ & & 3.5 & $ 4.4\pm1.3$ & $2.5^{+1.5}_{-1.9}$ & & 0.088 \\
AzTEC\_J100039.06+024128.6$^{b,c}$ & & 3.5 & $ 4.4\pm1.3$ & $2.5^{+1.4}_{-1.9}$ & & 0.089 \\
AzTEC\_J100012.42+022657.5 & & 3.5 & $ 4.5\pm1.3$ & $2.5^{+1.4}_{-2.1}$ & & 0.098 \\
\hline
AzTEC\_J100025.23+022608.0$^{a,d}$ & & 3.3 & $ 4.6\pm1.4$ & $1.9^{+1.2}_{-2.0}$ & & 0.144 \\
AzTEC\_J095939.01+022124.5$^{a,d}$ & & 3.2 & $ 6.5\pm2.0$ & $1.3^{+0.5}_{-1.7}$ & & 0.304 \\
\hline
\end{tabular}
\end{center}
Notes: $a)$ Sources have also been
detected with Bolocam (J. Aguirre, private communication); $b)$ AzTEC
sources with one or more candidate MIPS 24~\micron~counterpart
(\S~\ref{ssec:mips}); $c)$ AzTEC
sources with one or more candidate radio counterpart
(\S~\ref{ssec:radio}); $d)$ These sources have very non-Gaussian PFDs and
ill-defined local maxima due to low signal to noise. In these cases,
the de-boosted flux densities have been determined by the expectation
value of the flux given the PFD.
\end{table*}

\section{SIMULATIONS}
\label{sec:sim}
With the machinery described in \S~\ref{sec:dat} in place, it is
straightforward to determine various characteristics of our signal map
and our source identification process via Monte Carlo simulations.  We
generate synthetic source maps by populating our synthetic noise maps
with point source kernel-shaped sources.  Depending on the goal of the
simulation, sources of a given flux are randomly placed into the signal
or noise map one at a time, or entire populations of sources drawn from a
parametrised number-density distribution may be randomly distributed
(spatially) in a noise map.  When appropriate we determine characteristics
of our survey
with the former method in order to avoid biasing our results with the
(weak) prior of the input source distribution.

\subsection{\label{ssec:bayes}Flux De-Boosting}
Sources with low $S/N$ are detected at fluxes  
systematically higher than their intrinsic flux density 
when the source population 
increases in number with decreasing flux.  This well known but subtle
effect \citep[e.g.,][]{hogg98} 
becomes important when there are far more faint
sources, dimmer than the detection flux limit, than there are brighter
sources. In this instance it becomes
more likely that the numerous dim sources are boosted high by noise than 
the rarer bright sources are boosted to lower fluxes.  This is
particularly significant in surveys of SMGs, where detections are
almost always at low $S/N$ ($< 10$) 
and the intrinsic population is known to have a very steep luminosity
distribution \citep[e.g.,][and references therein]{scott06}.

For each source candidate we calculate a posterior flux distribution (PFD)
which describes the source's intrinsic flux in terms of probabilities.
The PFD
is calculated through an implementation of Bayes theorem similar to 
that used by \citet{coppin05,coppin06}. For an individual source
detected with measured flux density $S_m \pm \sigma_m$, the
probability distribution for its intrinsic flux density $S_i$ is given
by 
\be
p(S_i|S_m,\sigma_m) = {p(S_i)p(S_m,\sigma_m|S_i) \over p(S_m,\sigma_m)}
\ee
where $p(S_i)$ is the prior distribution of flux
densities, $p(S_m,\sigma_m|S_i)$ is the likelihood of observing the
data, and $p(S_m,\sigma_m)$ is a normalising constant. We assume a
Gaussian noise distribution for the likelihood of observing the
data, where
\be
\label{equ:gauss}
p(S_m,\sigma_m|S_i) = {1 \over \sqrt(2 \pi \sigma_m^2)} \mbox{exp}\left({-(S_m - S_i)^2 \over 2 \sigma_m^2}\right).
\ee
This assumption is justified by the Gaussian flux distribution observed in
jackknifed noise maps (thin line in Figure~\ref{fig:flux_hist}).
We use a Schechter function of the form:
\be
\label{equ:schechter}
{dN \over dS} = N^\prime \left({S \over S^\prime}\right)^{\alpha+1} \mbox{exp}(-S/S^\prime)
\ee
for the prior of the number counts, which we use to simulate the flux
distribution $p(S_i)$.
We adopt the best-fit parameters to the SCUBA
SHADES number counts \citep{coppin06}, scaled to 1.1~mm assuming an
850~\micron/1100~\micron~ spectral index of 2.7. The parameters for the
Schechter function prior are $N^\prime = 3200$~deg$^{-2}$~mJy$^{-1}$,
$S^\prime = 1.6$~mJy, and $\alpha = -2.0$. While the PFDs
will depend on the exact form of the source population,
we have verified that maximum likelihood flux densities derived from this
approach differ by less than 0.7~mJy (i.e. well within the
photometric error) for a variety of assumed models
(e.g. single
power law, Schechter function) and a wide range of parameters as
measured from previous SCUBA, Bolocam, and MAMBO SMG surveys
\citep[, respectively]{coppin06,laurent05,greve04}.

We estimate the prior distribution of flux densities by 
generating 10,000 noiseless sky realisations,
inserting sources with a uniform spatial distribution into a blank map
with the source population
described by Equation~\ref{equ:schechter}, where each source is
described by the point source kernel. The pixel histogram of flux
values from these sky maps gives an estimate of $p(S_i)$. 

\begin{figure}
\includegraphics[width=8cm]{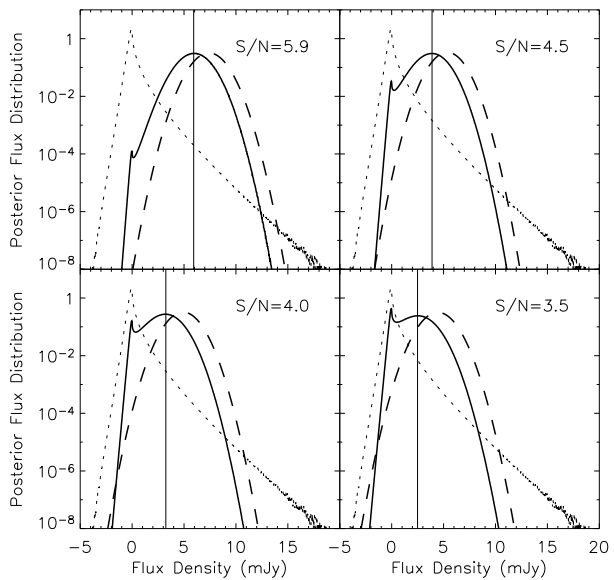}
\caption{Posterior flux distributions (PFDs) for a sample of four AzTEC
source candidates, whose $S/N$ values are representative of the range
observed in the entire source list. The dashed curve shows the
Gaussian distribution assumed for the measured source
flux distribution, $p(S_m,\sigma_m|S_i)$. The dotted curve is
$p(S_i)$, estimated from simulated sky maps as described in
\S~\ref{ssec:bayes}. The solid curve is the
posterior flux distribution, $p(S_i|S_m,\sigma_m)$. All
distributions have been normalised such that the integral under the
curve is equal to 1. The vertical line
indicates the local maximum of $p(S_i|S_m,\sigma_m)$, which gives the
de-boosted flux density of the source listed in column 5 of
Table~\ref{table:sources}.}
\label{fig:pfd} 
\end{figure}

A plot of the PFD for a sample of the AzTEC
source candidates is shown in Figure~\ref{fig:pfd}. These four sources
represent the range of measured fluxes in the catalogue and demonstrate
how the PFD varies according to the strength of the
detection. Strictly speaking, the PFD for a given source candidate depends
on both its detected flux \textit{and} noise, but this 
translates into a dependence on $S/N$ when the noise is uniform in the
map, which is approximately true in this case. 
We calculate the de-boosted flux density for
each source by locating the local maximum value of the PFD. These
values are listed in column 5 of
Table~\ref{table:sources}. The errors on the de-boosted fluxes shown in
Table~\ref{table:sources} represent the 68\% confidence interval.

Using the PFD, we estimate the probability that each detected source
candidate will be de-boosted to less than 0~mJy, which is listed in
column 7 of Table~\ref{table:sources} for each source
candidate. \citet{coppin05, coppin06} use these PFDs to exclude
source candidates that have $\ge$ 5\% probability of de-boosting to $<
0$ as a way to limit the source list to candidates which have a higher
probability of being real. While this may result in a source catalogue
with fewer false-detections, it could exclude many real sources
detected with low $S/N$ and reduce the completeness of the
source catalogue. Furthermore, while the de-boosted flux densities
derived from the PFDs are not very sensitive to the assumed source
population used to generate the prior distribution, 
the number of source candidates that meet the null threshold criterion
is sensitive to the exact form of the prior. For these reasons, we
choose to publish the entire list
of $\ge 3.5\sigma$ source candidates with the stipulation that some
fraction of this catalogue (in particular, source candidates with
$S/N < 4$) represent false-detections, as addressed in \S~\ref{ssec:fdr}.

\subsection{\label{ssec:fdr}False-Detection Rate}
Traditionally, a false-detection rate is the number of $>N\sigma$
peaks caused purely by noise and therefore appear at locations where
there are no real sources.  However, in surveys
such as ours, where the confused signal is significant relative to the
noise, every pixel in the map is affected by the presence of
sources.  Therefore, the definition
of false-detection rate becomes rather arbitrary.  Another
complication is that source confusion will increase the number of
positive and negative peaks in a map, beyond the number found in our
synthetic noise realisations.  A common practise is to quote the
false-detection rate as the number of \textit{negative} peaks detected
in the map
with $>N\sigma$ significance.  However, it is difficult to interpret
that number, mainly because source confusion may augment the number of
negative peaks differently from the number of positive peaks.

\begin{figure}
\includegraphics[width=8cm]{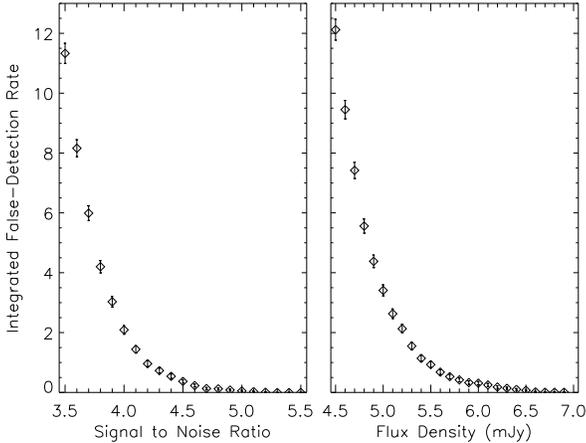}
\caption{Number of expected false-detections in the AzTEC/COSMOS
catalogue above a given signal to
noise (left) and measured source flux (right). The false-detection
rate determined here represents an upper
limit to the real number of false-detections that we expect (see
\S~\ref{ssec:fdr}).}
\label{fig:fdr} 
\end{figure}

Therefore, we show in Figure~\ref{fig:fdr} the number of ``sources''
detected when the usual source finding algorithm is applied to our
synthetic noise maps.  These curves are proportional to the number of
instances that a point with
zero flux in a {\em noiseless, beam-convolved} map of the sky is
detected above the given signal to noise ratio (or flux density).
Because nearly half the points
on a noiseless, beam-convolved map would have sub-zero flux (due to
AC-coupling), the curves of Figure~\ref{fig:fdr} give an upper limit to
the number of such sub-zero points that would spuriously be called
detections. Using this definition, the expected number of
false-detections for AzTEC/COSMOS sources with $S/N \ge 4.5$
is consistent with zero.

An alternative definition of false-detection rate could be
the number of ``source'' detections at points on the noiseless,
beam-convolved sky with intrinsic flux below $S$, where $S$
could be the detection threshold of a follow-up observation, for
instance with the SMA.  But we refrain from such speculation here because
the false-detection rate would depend on the source population as well
as the rather arbitrary $S$.

\subsection{\label{ssec:complete}Completeness}
\begin{figure}
\includegraphics[width=8cm]{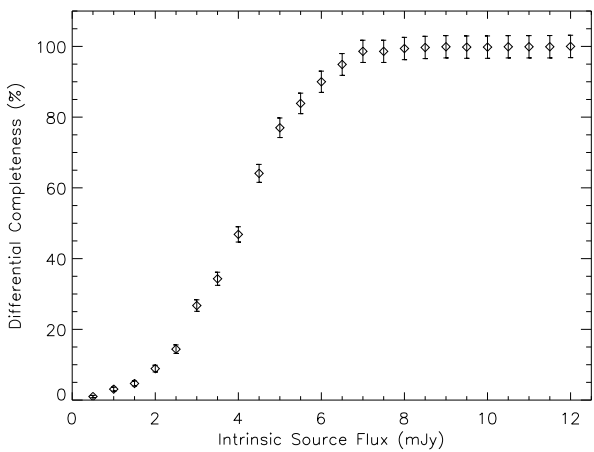}
\caption{Differential completeness versus intrinsic source flux density. The
errors are Poisson errors.}
\label{fig:completeness} 
\end{figure}

The differential completeness as a function of input source flux is
shown in Figure~\ref{fig:completeness}. Completeness is estimated by
injecting sources, one at a time, into the (sparsely populated) real 
signal map at random positions and checking if they are retrieved by our
standard source identification algorithm. Adding one source at a time
to the real signal map provides a valid estimate of the completeness
because it 1) accounts for the effects of random and confusion noise present in
the real map, 2) does not significantly alter the properties of the
real map (only one source input at a time), and 3) is not dependent on
a model of the source population (as is necessary for fully simulated
data-sets using noise maps). We inject a total of 1,000
sources per flux value, ranging from 0.5--12~mJy in steps of 0.5~mJy.
A source is considered to be recovered
if it is detected with $S/N \ge 3.5$ within
10\arcsec~of the input source position. We disregard any samples
where the input source is injected (or retrieved) within
10\arcsec~of a real $\ge 3.5\sigma$ source candidate in the map to
avoid confusion with real sources. The AzTEC/COSMOS survey is
50\% complete at 4~mJy, and 100\% complete at 7~mJy.

\subsection{\label{ssec:position_uncert}Positional Uncertainty}
\begin{figure}
\includegraphics[width=8cm]{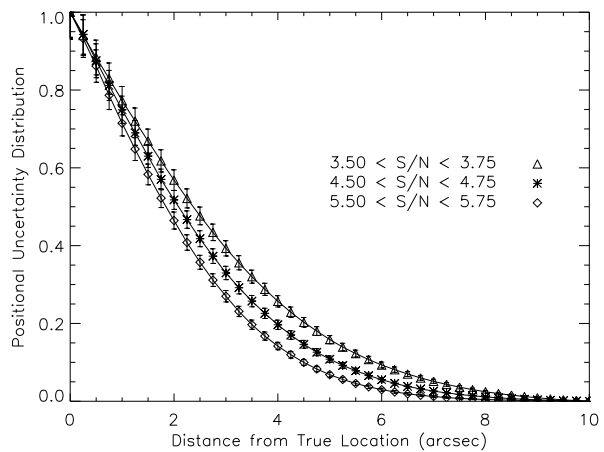}
\caption{The positional uncertainty distribution, $P(>D;S/N)$, for three sample
  signal to noise bins, showing the probability that an AzTEC source
  detected at a signal to noise ratio of $S/N$ will lie outside a
  distance $D$ from its true position.}
\label{fig:position} 
\end{figure}

The simulations described in \S~\ref{ssec:complete} offer a measure of
the error on the position of sources identified in the AzTEC map due
to the effects of both random and confusion noise. For the synthetic
sources that are recovered, we calculate the distance
between the input and output source positions and construct the
probability, $P(>D; S/N)$, that an AzTEC source with signal to noise
$S/N$ will be detected outside a distance $D$ of its true
position. This positional uncertainty measurement is not
sensitive to the contribution from positional errors arising from 
systematic and/or random errors introduced through corrections to the
pointing model (\S~\ref{ssec:pointing}). To account for this, we
assume Gaussian random pointing errors of 2\arcsec~in both RA and DEC
(see \S~\ref{ssec:radio} and Figure~\ref{fig:pointing}), and we
generate 100 random variates for
each recovered source to simulate pointing errors, which are
added to the measured output source position. A plot of the positional
uncertainty distribution as a function of distance for three different
signal to noise bins is shown in Figure~\ref{fig:position}. For all
$\ge 3.5\sigma$ AzTEC source candidates, the probability that an AzTEC
source will be detected within 4.5\arcsec~of its true position is
$\ge$ 80\%.

\section{COMPARISON WITH OTHER CATALOGUES}
\label{sec:multi_compare}
A detailed multi-wavelength study of AzTEC/COSMOS sources will be
deferred to Paper III. In this section, we discuss the 
confirmations of AzTEC sources with observations by Bolocam, identify potential
radio and MIPS~24~\micron~counterparts to the mm sources, and study the
faint mm emission from the rest of the radio/IR population.

\subsection{AzTEC Overlap with Bolocam Sources}
The AzTEC/COSMOS field overlaps slightly with the larger, shallower
Bolocam/COSMOS survey. Two of our high-significance
source candidates lie within 4\arcsec~of Bolocam-identified sources
detected with $S/N \ge 3.5$, confirming the reality of these
sources (J. Aguirre, private communication). The third Bolocam
source that lies within the AzTEC 75\% uniform coverage region is not
detected in our survey.

Two additional Bolocam-detected sources lie within the 25\%
uniform coverage region of the AzTEC map (the
2.5~mJy/beam contour shown in Figure~\ref{fig:weight_map}). We
tentatively confirm these two
Bolocam sources at the $\sim3\sigma$ level.
Though located 17--18\arcsec~from the Bolocam centroid, these AzTEC
source candidates are within the 95\% confidence radius of the
positional error in the Bolocam/COSMOS survey (J. Aguirre, private
communication). These four AzTEC sources which
are coincident with Bolocam detections are identified in
Table~\ref{table:sources}.

\subsection{\label{ssec:radio}The Corresponding Radio Population}

\begin{figure*}
\includegraphics[width=16cm]{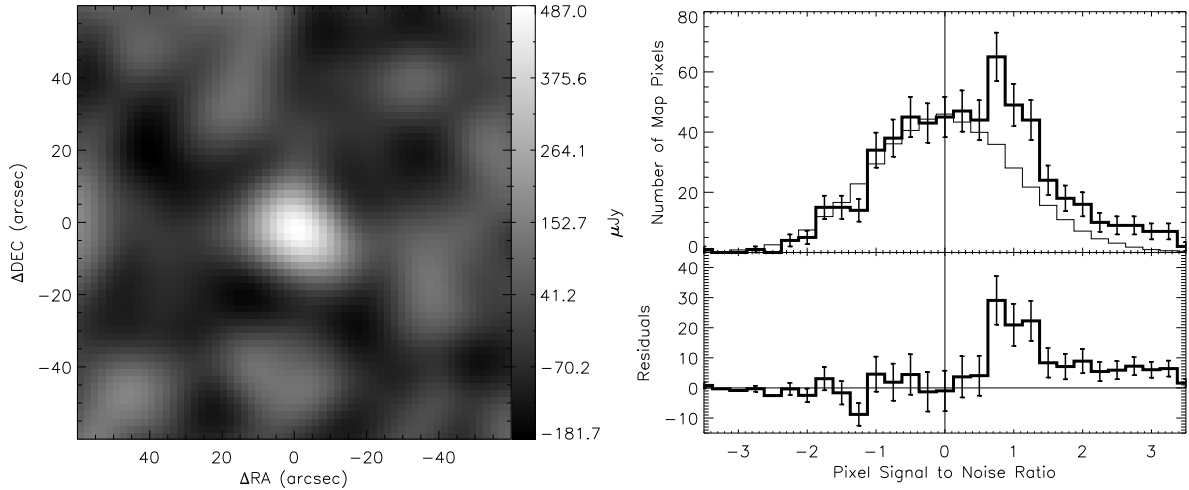}
\caption{\textbf{Left:} Average AzTEC map flux in
2\arcmin$\times$2\arcmin~cutouts centred at the 598 radio source
positions. We have excluded the positions of radio sources that are
located within 9\arcsec~of AzTEC peaks with $|S/N| \ge 3.5$. \textbf{Top
Right:} Histogram of the signal to noise ratio of the 1.1~mm map at the
radio source positions (thick line) versus that at positions chosen
randomly in the map (thin line). \textbf{Bottom Right:} The
difference between the two histograms above.}
\label{fig:stack_radio} 
\end{figure*}

The identification of radio counterparts has often been used to
improve on the positional uncertainty of SMGs
\citep[e.g.,][]{ivison02,ivison07,chapman03,chapman05,pope05,pope06}.  For 
this comparison we use the 4.5$\sigma$ catalogue from the
VLA/COSMOS survey \citep{schinnerer07}, which has a $1\sigma$ depth of
10.5~$\mu$Jy rms. To identify potential radio counterparts to our
mm-identified sources, we use a conservatively large search radius of
9\arcsec~from the measured AzTEC source position. If we assume that
the location of a candidate radio counterpart is the true location of
a given AzTEC source, then the probability that we detect the AzTEC
source at a distance greater than 9\arcsec~from the radio source is
given by the
positional uncertainty distribution that was calculated in
\S~\ref{ssec:position_uncert}, $P(>9\arcsec;S/N)$, which is $\le$ 1\%
for all $S/N$ values $\ge 3.5$. Thus using a search radius of
9\arcsec~makes it unlikely that we would fail to identify the radio
counterpart to an AzTEC source candidate, should it exist. On the
other hand, if the radio number density is high enough, we will expect
some fraction of false associations with AzTEC galaxies. 
We quantify this through the
``P-statistic'', which gives the probability that the first nearest
neighbour radio source will lie within a distance $\theta$ from a
given point and is given by
\be
  P(\theta) = 1 - e^{-n\pi\theta^2}
\ee
where $n$ is the number density of radio sources
\citep[e.g.,][]{scott89}. This
P-statistic is equivalent to the probability that a radio source will lie
within a distance $\theta$ of an AzTEC source candidate by chance. Assuming
uniform density (i.e. no clustering) of radio sources,
$n = 2350$~deg$^{-2}$ in this field, and thus $P(9\arcsec) = 4.5$\%. Hence 
we expect 4.5\% of radio sources identified within 9\arcsec~of an AzTEC
source candidate to be false associations.

For the list of source candidates in
Table~\ref{table:sources}, 15 have a single radio counterpart within
9\arcsec~of the AzTEC source position, and 3 have two radio sources
within 9\arcsec~of the AzTEC source position. AzTEC sources with at
least one candidate radio counterpart are identified in
Table~\ref{table:sources}. From the P-statistic, we expect
one of these 18 to be a false association. However, we may expect more
false associations than this if radio sources cluster on scales
smaller than 9\arcsec, making the local P-statistic in the
neighbourhood of mm sources higher. The fraction of AzTEC sources with
potential radio counterparts (36\%) is consistent with that found in
the SCUBA/SHADES survey \citep{ivison07} of 30--50\%, assuming the same
limiting flux (45~$\mu$Jy at 1.4~GHz), but is only marginally  
consistent (within 2$\sigma$, Poisson errors) with that of the
MAMBO/COSMOS survey
\citep{bertoldi07} of 67\%. Given the depth of the radio
survey from \citet[][7--8~$\mu$Jy]{bertoldi07}, this may
simply reflect
the relative completeness in the different radio catalogues. Our radio 
fraction could also be diluted by including low $S/N$ AzTEC sources,
which have a higher number of false-detections. The fraction of AzTEC
$\ge 4\sigma$ sources (only 2 false-detections expected) with
candidate radio counterparts is 12/27 (44\%) and agrees with the
\citet{bertoldi07} radio fraction within $1\sigma$.

We use the same radio catalogue to explore the weaker, confused
population of SMGs in the AzTEC map.  Figure~\ref{fig:stack_radio}
(left panel) shows the results of averaging the AzTEC map flux in
2\arcmin$\times$2\arcmin~postage stamps extracted from regions
centred at the 598 radio source positions that lie within the AzTEC
map boundary. Since we
compute a weighted average for each pixel, we extend this analysis
to the noisier edges of the mm map (10\% coverage region, with an area
of 0.28~deg$^2$).  All
radio sources that have candidate AzTEC counterparts detected at
$\ge 3.5\sigma$ or $\le -3.5\sigma$ have been excluded in order to
restrict this analysis to radio sources with faint AzTEC emission,
below the $S/N$ threshold used for discrete source identification.
The 8.06$\sigma$ stacked signal implies a
mean 1.1~mm flux of 487 $\pm$ 60~$\mu$Jy for the radio sources in the
catalogue. No significant difference in the average 1.1~mm flux is detected
when we stack separately on two groups of radio sources divided by
their 1.4~GHz flux. For radio sources with flux density
$> 66~\mu$Jy (293/598), the stacked 1.1~mm signal is
530 $\pm$ 87~$\mu$Jy, while the stacked 1.1~mm flux
for radio sources $\le 66~\mu$Jy (305/598) is
465 $\pm$ 84~$\mu$Jy. These values differ by only 13\%
and agree within the errors.

In the top right panel of Figure~\ref{fig:stack_radio}, we show a
histogram of the 1.1~mm $S/N$ ratio at the
location of all 598 radio sources. For comparison, we generate 100
fake catalogues, each with 598 positions chosen randomly across the
AzTEC map, and construct the histogram of AzTEC $S/N$ ratio at these
locations. Since these positions were chosen at random, we expect that
the distribution of $S/N$ values should be nearly symmetric about zero. The
bottom right panel of Figure~\ref{fig:stack_radio} shows the
difference between the histogram of the $S/N$ ratios at the radio source
positions and that at the random positions. This clearly demonstrates
that there is a significant contribution to the stacked flux image
from low $S/N$ mm sources. Roughly 1/2 of the stacked signal arises 
from sources with $S/N < 1.8$ that fall below the detection threshold
for source identification. This analysis demonstrates
that the AzTEC map is sensitive to very faint millimetre emission
down to flux levels on order of the 1$\sigma$ rms of the map.

\begin{figure}
\includegraphics[width=8cm]{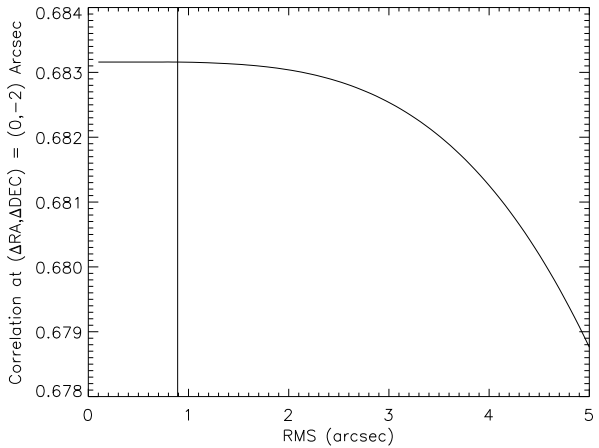}
\caption{Amplitude of the cross-correlation between the map 
         in Figure~\ref{fig:stack_radio} and a map constructed by
         convolving the point source kernel with a Gaussian with
         standard deviation $\sigma_p$. For all values of $\sigma_p$,
         the maximum correlation occurs at ($\Delta$RA, $\Delta$DEC) =
         (0\arcsec, -2\arcsec).}
\label{fig:pointing} 
\end{figure}

The radio source stacking analysis can also be used to estimate the
residual systematic and rms pointing errors in the AzTEC map due to
errors in the astrometry.  The stacked signal peaks at
($\Delta$RA, $\Delta$DEC)$ = (0.4\arcsec, -2.1\arcsec)$, indicating a
potential small systematic offset.
Noise in the pointing solution leads to a broadening of the stacked
signal, and so we use a measure of this broadening to determine the rms
pointing uncertainty of our AzTEC observations.  The model is as
follows: assuming that the pointing errors are random and Gaussian
distributed with mean zero and standard deviation $\sigma_p$, the
stacked AzTEC flux should be equal to the convolution of a Gaussian
(with standard deviation $\sigma_p$) with the point source kernel. We
calculate the cross-correlation of the stacked AzTEC flux at the radio
source locations with this model, varying $\sigma_p$. We find that for
all values of $\sigma_p$, the maximum value of the cross-correlation
matrix is at an offset of zero in RA and -2\arcsec~in DEC, consistent
with a small systematic pointing offset. Figure~\ref{fig:pointing}
shows the value of the maximum correlation as a function of pointing
uncertainty, $\sigma_p$.  The strongest correlation occurs for
$\sigma_p = 0.89$\arcsec. However, the curve becomes very flat at
$\sigma_p < 2$\arcsec~because the stacked image itself is limited to
2\arcsec~pixelization. Also, if radio sources in the COSMOS field
cluster on scales $<$ 2\arcsec, this would also broaden the width of the
stacked signal, further complicating this estimate. Though we cannot
accurately measure the value of
$\sigma_p$ with this technique when $\sigma_p$ is small, we can state
with confidence that $\sigma_p <$ 2\arcsec, and we adopt this as a
conservative estimate of the error in the astrometry in our map. We
combine this error with the measured distances between input
and output source positions as described in
\S~\ref{ssec:position_uncert} to determine the positional uncertainty
distribution shown in Figure~\ref{fig:position}.

\subsection{\label{ssec:mips}Coincident 24~\micron~Detections}

\begin{table*}
\begin{center}
\begin{tabular}{lcccc}
\hline
          & AzTEC source candidates & AzTEC source candidates &               &  Catalogue\\
          & with $\ge$1 counterpart & with 2 counterparts     & $P(9\arcsec)$ &  Completeness\\
\hline
Radio     & 18/50 (36\%)                      & 3/50 (6\%)                                    & 4.5\%                       &  4$\sigma$=45$\mu$Jy\\
24 $\mu$m & 32/49 (65\%)                      & 2/49 (4\%)                                    & 24.5\%                      &  5$\sigma$=60$\mu$Jy\\
\hline
\end{tabular}
\end{center}
\caption{\label{table:radiomips} Comparison of AzTEC source candidates
with radio and MIPS 24~\micron~sources, using a search radius of 9\arcsec.
$P(9\arcsec)$ is the probability of a chance coincidence within the 
9\arcsec~search radius.}
\end{table*}

\begin{figure*}
\includegraphics[width=16cm]{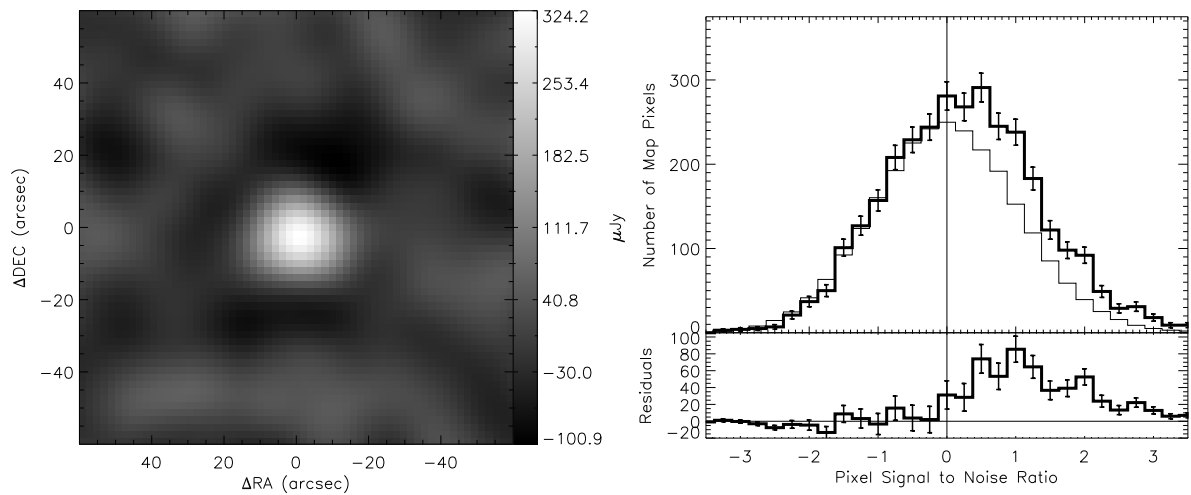}
\caption{\textbf{Left:} Average AzTEC map flux in
2\arcmin$\times$2\arcmin~cutouts centred at 3129 24~\micron~source
positions. We have excluded the positions of 24~\micron~sources that are
located within 9\arcsec~of AzTEC peaks with $|S/N| \ge 3.5$. \textbf{Top
Right:} Histogram of the signal to noise ratio of the 1.1~mm map at the
24~\micron~source positions (thick line) versus that at positions chosen
randomly in the map (thin line). \textbf{Bottom Right:} The
difference between the two histograms above.}
\label{fig:stack_mips} 
\end{figure*}

A similar comparison can be made to sources detected at
24~\micron~by the \textit{Spitzer}/MIPS instrument in the COSMOS deep survey
\citep{sanders07}. There are 2082 24~\micron~sources with
$S/N \ge 5$ ($S_{24\micron} \ge 60~\mu$Jy) within the 75\% uniform
coverage region of the AzTEC/COSMOS map, and 49/50 AzTEC source
candidates within the coverage of the MIPS~24~\micron~image.  Of
these, 30 individual 24~\micron~sources are found within 9\arcsec~of
an AzTEC source, while 2 AzTEC sources have two 24~\micron~sources
within a 9\arcsec~radius.  AzTEC sources with one or more
potential MIPS 24~\micron~counterparts are identified in
Table~\ref{table:sources}. The source density of 24~\micron~sources in
this field is quite large (14280~deg$^{-2}$) and the probability of
chance coincidence within 9\arcsec~is 24.5\%, so we expect 12 false
associations. As shown in \citet{younger07}, it is not uncommon to
find an unrelated 24~\micron~source within 9\arcsec~of an SMG. We therefore do
not use the 24~\micron~catalogue as a signpost for
mm-wavelength emission. A summary of the number of AzTEC source
candidates with potential radio and 24~\micron~counterparts is given
in Table~\ref{table:radiomips}. A detailed multi-wavelength study of
the AzTEC sources in this field will be presented in Paper III.

We perform the same stacking analysis as done for the radio catalogue on
the 24~\micron~catalogue.  The results are shown in
Figure~\ref{fig:stack_mips}.  Again, MIPS sources within 9\arcsec~of
an AzTEC pixel with $S_{\rm{1.1mm}} \ge 3.5\sigma$ or
$S_{\rm{1.1mm}} \le -3.5\sigma$ have been excluded.  This leaves 3129 MIPS
sources within
the extended AzTEC map.  The stacked signal strength is 12.8$\sigma$, and the
mean 1.1~mm flux of these sources is 324 $\pm$ 25~$\mu$Jy. A histogram of
the 1.1~mm $S/N$ ratio at the location of all 3129 MIPS 24~\micron~sources
is shown in the right panel of Figure~\ref{fig:stack_mips}, demonstrating
that the stacked signal is dominated by low ($< 2\sigma$) $S/N$ mm sources.

\section{The Contribution of AzTEC Sources in COSMOS to the Cosmic
Infrared Background}
\label{sec:cib}

Using the de-boosted 1.1~mm AzTEC flux densities derived from the
PFDs, we sum the flux densities of the $\ge 3.5\sigma$ source
candidates to determine the resolved fraction of the Cosmic Infrared
Background (CIB) in this survey.  An integrated flux of
1.3~Jy~deg$^{-2}$ from those galaxies in the AzTEC catalogue
(Table~\ref{table:sources}) is compared to 18--24~Jy~deg$^{-2}$ from
the CIB measured by \textit{COBE}-FIRAS at 1.1~mm
\citep{puget96, fixsen98}, demonstrating that we have resolved 
5.3--7.1\% of the CIB into bright millimetre-wavelength sources in the 
COSMOS field.
This value is an overestimate of the real CIB resolved in this study
because at least some of the source candidates are false-detections
(random noise peaks). Also, there appears to be an over-density
of bright mm sources in this field, in which case the local CIB
would be larger than the average value measured in \citet{puget96} and
\citet{fixsen98}.

Furthermore, we can estimate the fraction of the millimetre CIB
resolved by the entire radio population in the COSMOS field. Using the
stacked analysis described in \S~\ref{ssec:radio}, we first calculate the
average millimetre flux of the faint AzTEC sources with $S/N < 3.5$
that are associated with the 598 radio counterparts distributed over
0.28~deg$^2$, which is 487 $\pm$ 60~$\mu$Jy at 1.1~mm, or
1.0 $\pm$ 0.1~Jy~deg$^{-2}$.  This
resolved fraction of 4.3--5.7\% of the millimeter CIB is comparable to
that measured from stacking the 850~\micron~flux at the position of
1.4~GHz radio sources in the SCUBA/GOODS-N field, where \citet{wang06}
resolve 3.4--4.8\% of the CIB (excluding the contribution from
$\ge 4\sigma$ sources) using a radio catalogue with a similar limiting
flux (40~$\mu$Jy) as the COSMOS radio catalogue.  Next we add 
the contribution of 0.46~Jy~deg$^2$ at 1.1~mm from
the 18 bright ($S/N \ge 3.5$) AzTEC sources in
Table~\ref{table:sources} that have
radio counterparts. We therefore conclude that our AzTEC map has
resolved a total 1.1~mm flux of 1.46~deg$^2$, or 7 $\pm$ 1\% of
the CIB, due to the full population of radio sources in COSMOS.

Finally, considering the average millimetre flux of the faint
population ($< 3.5\sigma$) of AzTEC galaxies at the positions of the MIPS
24~\micron~sources (\S~\ref{ssec:mips}), we estimate a total 1.1~mm flux of
4.4 $\pm$ 0.3~Jy~deg$^{-2}$, thereby resolving 18.3--24.4\% of the
CIB. Similarly \citet{wang06} resolve 13.4--19.0\% of the CIB from
their 850~\micron~stacking analysis of MIPS~24~\micron~ sources in the
SCUBA/GOODS-N map. Although their 24~\micron~catalogue is slightly
shallower than the COSMOS MIPS 24~\micron~source catalogue (80~$\mu$Jy
and 60~$\mu$Jy, respectively), these CIB fractions agree within the
errors of the measurements.

\section{CONCLUSIONS}
\label{sec:con} 
We have imaged a 0.15~deg$^2$ region within the COSMOS field with
AzTEC, a new mm-wavelength camera, with uniform sensitivity of 1.3~mJy/beam at
1.1~mm. We have identified 50 source candidates in the
AzTEC/COSMOS map with signal to noise ratio $\ge$ 3.5, 16 of
which are detected with $S/N \ge 4.5$, where the expected number
of false-detections is zero. Seven of the $\ge 5\sigma$ source
candidates have been followed up and
confirmed with SMA imaging \citep{younger07}. The 
sources are spread throughout the field, with only 3 located in the
$z = 0.73$ cluster environment. Our catalogue is
50\% complete at an intrinsic flux density of 4~mJy, and is 100\%
complete at 7~mJy.  The positional uncertainty of these AzTEC sources
due to random and confusion noise is determined through simulations
which show that sources with $S/N \ge 3.5$ have $\ge$ 80\% probability of
being detected within 4.5\arcsec of their true location.

Comparing our
$\ge 3.5\sigma$ source candidate list with the radio source catalogue of
\citet{schinnerer07}, we find that the fraction of AzTEC sources with
potential radio counterparts is 36\% and is consistent with that found
in the SCUBA/SHADES survey \citep{ivison07} at similar flux
levels. From averaging the AzTEC map flux at the locations of the
radio and MIPS~24~\micron~\citep{sanders07} source positions, we
statistically detect the faint mm emission (below our detection
threshold) of radio and MIPS~24~\micron~sources and thereby
demonstrate that errors in the mean astrometry of our map arising from
the pointing model are small ($<2$\arcsec). Estimates of the resolved
fraction of the millimetre CIB due to these radio and mid-IR galaxy
populations is 7 $\pm$ 1\% and 21 $\pm$ 3\% respectively.

The AzTEC/COSMOS field samples a region of high galaxy over-density
compared to the regions imaged with MAMBO and Bolocam, and our
AzTEC/COSMOS map contains a large number of very bright mm sources (9
with corrected flux density $\ge 5$~mJy, where 4--5 are expected for an
unbiased field). We will present a complete analysis of the
relationship between the SMG population and the foreground galaxy
population in Paper II of this series.

The availability of extensive high quality multi-wavelength data from
the radio to the X-ray makes the follow-up analysis of the detected
sources readily possible and will allow us to study the nature of
these sources. A full analysis of the multi-wavelength properties of
the sources detected in this survey will be presented in Paper III.


\section*{acknowledgements}

The authors are grateful to J. Aguirre, J. Karakla, K. Souccar,
I. Coulson, R. Tilanus,
R. Kackley, D. Haig, S. Doyle, J. Lowenthal, and the observatory staff
at the JCMT who made these
observations possible.  Support for this work was provided in part by
the NSF grant AST 05-40852 and the grant from the Korea Science \&
Engineering Foundation (KOSEF) under a cooperative Astrophysical
Research Center of the Structure and Evolution of the Cosmos
(ARCSEC). IA and DHH acknowledge partial support by
CONACyT from research grants 39953-F and 39548-F.
This research has made use of the NASA/IPAC Extragalactic Database
(NED) which is operated by the Jet Propulsion Laboratory, California
Institute of Technology, under contract with the National Aeronautics
and Space Administration. 


\bibliography{references}

\end{document}